\documentclass[%
aps,
prl,
twocolumn,
superscriptaddress,
amsmath,amssymb,
floatfix,
]{revtex4-2}

\usepackage{graphicx}
\usepackage{dcolumn}
\usepackage{bm}
\usepackage{orcidlink} 
\usepackage{float}
\usepackage{comment} 
\usepackage[mathlines]{lineno}
\usepackage{hyperref}
\hypersetup{hypertexnames=false}
\makeatletter
\renewcommand\paragraph{\@startsection{paragraph}{4}{\parindent}%
  {2ex \@plus1ex \@minus.2ex}{-0.01em}{\normalfont\normalsize\itshape}}
\makeatother

\begin{document}
\preprint{APS/123-QED}
\title{Constraints on Solar Reflected Dark Matter from XENON1T and XENONnT Data}
\newcommand{\bologna}{\affiliation{Department of Physics and Astronomy, University of Bologna and INFN-Bologna, 40126 Bologna, Italy}}
\newcommand{\chicago}{\affiliation{Department of Physics, Enrico Fermi Institute \& Kavli Institute for Cosmological Physics, University of Chicago, Chicago, IL 60637, USA}}
\newcommand{\coimbra}{\affiliation{LIBPhys, Department of Physics, University of Coimbra, 3004-516 Coimbra, Portugal}}
\newcommand{\columbia}{\affiliation{Physics Department, Columbia University, New York, NY 10027, USA}}
\newcommand{\lngs}{\affiliation{INFN-Laboratori Nazionali del Gran Sasso and Gran Sasso Science Institute, 67100 L'Aquila, Italy}}
\newcommand{\mainz}{\affiliation{Institut f\"ur Physik \& Exzellenzcluster PRISMA$^{+}$, Johannes Gutenberg-Universit\"at Mainz, 55099 Mainz, Germany}}
\newcommand{\mpik}{\affiliation{Max-Planck-Institut f\"ur Kernphysik, 69117 Heidelberg, Germany}}
\newcommand{\munster}{\affiliation{Institut f\"ur Kernphysik, University of M\"unster, 48149 M\"unster, Germany}}
\newcommand{\nikhef}{\affiliation{Nikhef and the University of Amsterdam, Science Park, 1098XG Amsterdam, Netherlands}}
\newcommand{\nyuad}{\affiliation{New York University Abu Dhabi - Center for Astro, Particle and Planetary Physics, Abu Dhabi, United Arab Emirates}}
\newcommand{\purdue}{\affiliation{Department of Physics and Astronomy, Purdue University, West Lafayette, IN 47907, USA}}
\newcommand{\rice}{\affiliation{Department of Physics and Astronomy, Rice University, Houston, TX 77005, USA}}
\newcommand{\stockholm}{\affiliation{Oskar Klein Centre, Department of Physics, Stockholm University, AlbaNova, Stockholm SE-10691, Sweden}}
\newcommand{\subatech}{\affiliation{SUBATECH, IMT Atlantique, CNRS/IN2P3, Nantes Universit\'e, Nantes 44307, France}}
\newcommand{\torino}{\affiliation{INAF-Astrophysical Observatory of Torino, Department of Physics, University  of  Torino and  INFN-Torino,  10125  Torino,  Italy}}
\newcommand{\ucsd}{\affiliation{Department of Physics, University of California San Diego, La Jolla, CA 92093, USA}}
\newcommand{\wis}{\affiliation{Department of Particle Physics and Astrophysics, Weizmann Institute of Science, Rehovot 7610001, Israel}}
\newcommand{\zurich}{\affiliation{Physik-Institut, University of Z\"urich, 8057  Z\"urich, Switzerland}}
\newcommand{\paris}{\affiliation{LPNHE, Sorbonne Universit\'{e}, CNRS/IN2P3, 75005 Paris, France}}
\newcommand{\freiburg}{\affiliation{Physikalisches Institut, Universit\"at Freiburg, 79104 Freiburg, Germany}}
\newcommand{\napels}{\affiliation{Department of Physics ``Ettore Pancini'', University of Napoli and INFN-Napoli, 80126 Napoli, Italy}}
\newcommand{\nagoya}{\affiliation{Kobayashi-Maskawa Institute for the Origin of Particles and the Universe, and Institute for Space-Earth Environmental Research, Nagoya University, Furo-cho, Chikusa-ku, Nagoya, Aichi 464-8602, Japan}}
\newcommand{\laquila}{\affiliation{Department of Physics and Chemistry, University of L'Aquila, 67100 L'Aquila, Italy}}
\newcommand{\tokyo}{\affiliation{Kamioka Observatory, Institute for Cosmic Ray Research, and Kavli Institute for the Physics and Mathematics of the Universe (WPI), University of Tokyo, Higashi-Mozumi, Kamioka, Hida, Gifu 506-1205, Japan}}
\newcommand{\kobe}{\affiliation{Department of Physics, Kobe University, Kobe, Hyogo 657-8501, Japan}}
\newcommand{\kit}{\affiliation{Institute for Astroparticle Physics, Karlsruhe Institute of Technology, 76021 Karlsruhe, Germany}}
\newcommand{\tsinghua}{\affiliation{Department of Physics \& Center for High Energy Physics, Tsinghua University, Beijing 100084, P.R. China}}
\newcommand{\ferrara}{\affiliation{INFN-Ferrara and Dip. di Fisica e Scienze della Terra, Universit\`a di Ferrara, 44122 Ferrara, Italy}}
\newcommand{\groningen}{\affiliation{Nikhef and the University of Groningen, Van Swinderen Institute, 9747AG Groningen, Netherlands}}
\newcommand{\westlake}{\affiliation{Department of Physics, School of Science, Westlake University, Hangzhou 310030, P.R. China}}
\newcommand{\shenzhen}{\affiliation{School of Science and Engineering, The Chinese University of Hong Kong (Shenzhen), Shenzhen, Guangdong, 518172, P.R. China}}
\newcommand{\coimbrapoli}{\affiliation{Coimbra Polytechnic - ISEC, 3030-199 Coimbra, Portugal}}
\newcommand{\uniheidelberg}{\affiliation{Physikalisches Institut, Universit\"at Heidelberg, Heidelberg, Germany}}
\newcommand{\roma}{\affiliation{INFN-Roma Tre, 00146 Roma, Italy}}
\newcommand{\bucknell}{\affiliation{Department of Physics \& Astronomy, Bucknell University, Lewisburg, PA, USA}}
\author{E.~Aprile\,\orcidlink{0000-0001-6595-7098}}\columbia
\author{J.~Aalbers\,\orcidlink{0000-0003-0030-0030}}\groningen
\author{K.~Abe\,\orcidlink{0009-0000-9620-788X}}\tokyo
\author{M.~Adrover\,\orcidlink{0123-4567-8901-2345}}\zurich
\author{S.~Ahmed Maouloud\,\orcidlink{0000-0002-0844-4576}}\paris
\author{L.~Althueser\,\orcidlink{0000-0002-5468-4298}}\munster
\author{B.~Andrieu\,\orcidlink{0009-0002-6485-4163}}\paris
\author{E.~Angelino\,\orcidlink{0000-0002-6695-4355}}\lngs
\author{D.~Ant\'on~Martin\,\orcidlink{0000-0001-7725-5552}}\chicago
\author{S.~R.~Armbruster\,\orcidlink{0009-0009-6440-1210}}\mpik
% ...
\collaboration{XENON Collaboration}\email[]{xenon@lngs.infn.it}\noaffiliation

\begin{abstract}
We report on a search for sub-GeV dark matter upscattered via the solar reflection mechanism in the heavy mediator scenario. Under the standard halo model (SHM), keV--MeV dark matter produces nuclear recoils with energies below the detection threshold of liquid xenon time projection chambers. We enhance sensitivity to low-mass dark matter (DM) by considering dark matter-electron scattering, employing dedicated event selections to reduce the detection threshold, and exploiting the additional kinetic energy imparted to the dark matter particle by solar upscattering. Using XENON1T ionization-only and XENONnT low-energy electronic recoil datasets, we exclude previously unconstrained DM-electron scattering cross section for masses between $4.6$ and $20\, \text{keV/}c^2$, and between $0.2$ and $2\, \text{MeV/}c^2$, reaching a minimum of $3.41\times10^{-39}\, \text{cm}^2$ for a mass of $0.3\, \text{MeV/}c^2$ at 90\% confidence level.
\end{abstract}

\maketitle

\paragraph{\label{sec:intro}Introduction\textemdash}
A wide range of astrophysical and cosmological observations point to the existence of dark matter (DM), a nonluminous component that dominates the matter content of the Universe. These observations are most naturally explained by one or more new particle species beyond the standard model~\cite{Bertone:2004pz}. Despite extensive searches, DM particles remain undetected~\cite{Roszkowski:2017nbc}. Among direct detection efforts, liquid xenon time projection chambers (TPCs) currently place the most stringent limits on DM-nucleon interactions in the GeV--TeV mass range~\cite{XENON:2025vwd,LZ:2024zvo,PandaX:2024qfu}.

The XENONnT experiment, located at the INFN Laboratori Nazionali del Gran Sasso, is the latest iteration of the multistage XENON program, succeeding XENON1T~\cite{XENON:2017lvq}. It consists of three nested detectors: a water Cherenkov muon veto~\cite{XENON1T:2014eqx}, a water Cherenkov neutron veto to suppress radiogenic neutron background~\cite{XENON:2024fxf}, and a double-walled cryostat containing $8.5$ metric tons of liquid xenon. Within the cryostat lies a dual-phase TPC with $5.9$ metric tons of instrumented liquid xenon in a cylindrical volume ($1.49\, \text{m}$ height, $1.33\, \text{m}$ diameter) with a thin gaseous xenon layer on the liquid xenon~\cite{nt_analysis_paper1,XENON:2024wpa}. Several key systems, such as the cryogenic gas purification and krypton distillation column, were developed for XENON1T. Notable improvements in the XENONnT experiment include the liquid xenon purification~\cite{Plante:2022khm} and high-flow radon removal systems~\cite{Murra:2022mlr}.

When a particle interacts with xenon atoms in the TPC, it deposits energy, creating prompt scintillation photons and ionization electrons. The prompt scintillation photons are detected by the arrays of photomultiplier tubes (PMTs) at the top and bottom of the TPC, forming the prompt scintillation signal (S1)~\cite{FUJII2015293}. The electrons drift upward under an electric field generated by the cathode at the bottom of the TPC and the gate electrode near the liquid surface. Upon reaching the liquid-gas interface, the electrons are extracted into the gaseous xenon by a stronger extraction field~\cite{XENON:2024wpa}. The accelerated electrons produce further electroluminescence in the gaseous xenon, resulting in the delayed ionization signal (S2) which is also detected by the PMT arrays~\cite{lansiart_seigneur_moretti_morucci_1976}.

In the SHM, DM particles in the local Galactic DM halo have speeds up to $\sim800\, \text{km/s}$ in Earth's laboratory frame~\cite{Evans:2018bqy}. For sub-GeV DM, such velocities typically produce recoil energies below the detection thresholds of liquid xenon TPCs. These thresholds are $1\, \text{keV}$ for combined scintillation and ionization signals~\cite{PhysRevLett.129.161805} and $13.7\, \text{eV}$ for the average ionization-only signals~\cite{XENON:2024znc}, restricting sensitivity to GeV--TeV and MeV masses respectively.

To extend the sensitivity to light DM, we make the following three analysis choices. First, we consider DM-electron scattering, which allows detection of lighter DM compared to DM-nucleus interactions. Second, we use specialized event selections considering ionization-only signals (S2-only) and low-energy electronic recoil (ER) signals to enhance sensitivity to smaller energy deposits. Finally, DM particles can gain kinetic energy by scattering with astrophysical targets~\cite{Bringmann:2018cvk,Wang:2021jic,Das:2021lcr,Emken:2017hnp,An:2017ojc}. We consider upscattering by the Sun~\cite{Emken:2017hnp,An:2017ojc}, which can reflect halo DM toward Earth with increased kinetic energy and hence higher detection efficiency.

In this Letter, we examine the upscattering of leptophilic sub-GeV DM particles in the Galactic halo by electrons in the solar plasma, also known as solar reflected dark matter (SRDM). The upscattering is assumed to occur exclusively with electrons, with no interactions with hadrons~\cite{Emken:2021lgc,Kopp:2009et}. We focus on the scenario in which the DM-electron interaction is mediated by a heavy particle, resulting in a contact interaction, which leads to a momentum-independent cross section and DM form factor of $F_\text{DM}(q)=1$~\cite{Emken:2021lgc}.

The DM-electron interaction responsible for upscattering DM in the Sun is the same interaction that produces ER signals in the detector, so no additional coupling is assumed. Using data from the XENON1T S2-only~\cite{PhysRevLett.123.251801} and XENONnT low-energy ER~\cite{PhysRevLett.129.161805} analyses, we derive 90\% confidence level (CL) confidence intervals on the DM-electron cross section for DM masses between $4.6\, \text{keV/}c^2$ and $9\, \text{MeV/}c^2$.

\paragraph{\label{sec:SRDM}Solar reflected dark matter\textemdash}
The final kinetic energy of an upscattered DM particle has an analytic expression if the DM particle only scattered once in the solar plasma~\cite{Emken:2017hnp}. However, for many combinations of the DM mass and DM-electron cross section, multiple scattering becomes significant. This renders the analytic treatment intractable and the SRDM differential flux must be evaluated numerically. In this Letter, we compute the SRDM differential flux using Monte Carlo simulations implemented in DaMaSCUS-SUN~\cite{DaMaSCUSsun}.

The computation of the SRDM flux follows the simulation procedure outlined in~\cite{Emken:2021lgc}. The process begins by sampling the initial position and velocity of a DM particle falling into the Sun and propagating the particle through the solar interior. As the DM particle traverses the solar plasma, it may scatter off electrons according to the assumed DM-electron scattering cross section. During its passage, the DM particle may undergo one of three processes: traverse the Sun without scattering, become gravitationally captured through repeated interactions, or scatter at least once and escape, thus contributing to the SRDM population.

The probability of solar reflection is estimated as the fraction of simulated DM particles that scatter at least once and subsequently escape, relative to the total number of simulated trajectories. The reflected DM particles are then propagated to Earth, where their velocity distribution is recorded. The resulting SRDM differential flux on Earth in units of $\mathrm{(km/s)^{-1}\,cm^{-2}\,s^{-1}}$ is given by
\begin{equation}
    \begin{aligned}
        \frac{d\Phi_\odot}{dv_\chi} &=\frac{1}{4\pi l^2}\frac{N_\text{refl}}{N_\text{sim}}\Gamma(m_\chi)f_\odot(v_\chi)\, ,
    \end{aligned}
    \label{eq:srdm_mc_diff_flux}
\end{equation}
where $l=1\, \text{A.U.}$ is the distance between Earth and the Sun, $N_\text{sim}=10^5$ is the total number of simulated DM trajectories, $N_\text{refl}$ is the number of simulated events with at least one scattering in the Sun, $\Gamma(m_\chi)$ represents the total rate of halo DM particles of mass $m_\chi$ falling into the Sun in the simulation, and $f_\odot(v_\chi)$ denotes the probability density function of the SRDM speed on Earth $v_\chi$, as obtained from simulation.

The predicted DM-electron scattering rates depend on the choice of atomic form factors, which affects both the SRDM and standard halo DM analyses. In this Letter, we use DaMaSCUS-SUN, which employs xenon atomic response functions from Ref.~\cite{Catena:2019gfa}. Comparisons among atomic calculations in the literature find differences at the level of $\mathcal{O}(10\%)$ in the relevant energy range~\cite{Pandey:2018esq,Liu:2024bhy}, small compared to the span of the final upper limits. We adopt the standard solar model (AGSS09) for this analysis~\cite{Serenelli:2009yc,patscott_sunmodel}. Although various solar models exist in the literature, the solar reflection results are only weakly sensitive to this choice~\cite{Emken:2021lgc}.

We neglect attenuation of the SRDM flux due to Earth's overburden as it is negligible at the DM-electron cross sections considered in this Letter. The extent of attenuation is model dependent, with nuclear recoil (NR) interactions producing stronger suppression than ER interactions. Even under the conservative assumption of equal contributions from ER and NR interactions, the critical DM-electron cross section above which Earth’s attenuation matters is $\sim10^{-31}\, \text{cm}^2$ for a $10\, \text{MeV/}c^2$ DM. This critical DM-electron cross section is larger for lighter DM~\cite{Crisler:2018gci}. In this Letter, we consider DM masses up to $9\, \text{MeV/}c^2$ and cross sections in the $10^{-39}$ to $10^{-35}\, \text{cm}^2$ range, where overburden effects are safely negligible.

\begin{figure}[ht]
    \centering
    \includegraphics[width=\columnwidth]{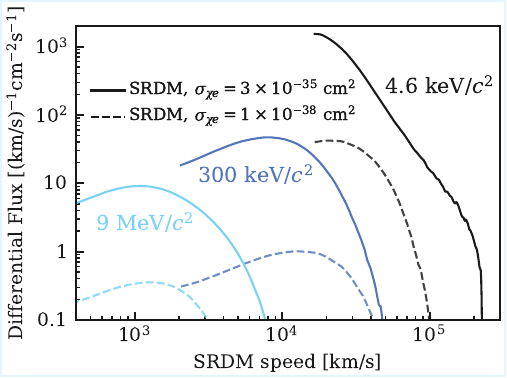}
    \caption{SRDM differential flux in a terrestrial detector as a function of SRDM speed, computed via Monte Carlo simulations~\cite{DaMaSCUSsun}. The shape of the differential flux depends on the DM mass (indicated by colors) and also the DM-electron cross section (indicated by line styles). The differential fluxes are computed to the lowest DM speed that can produce detectable signals in the detector for the given SRDM mass.}
    \label{fig1:diff_flux}
\end{figure}

The resulting SRDM differential flux observed in a terrestrial detector is shown in Fig.~\ref{fig1:diff_flux} as a function of DM speed in Earth's laboratory frame.
A larger DM-electron scattering cross section generally leads to a higher interaction rate in the detector, as DM is more likely to scatter with electrons in the liquid xenon target. It also increases the probability of DM reflection in the Sun, increasing the boost in velocity the DM particle receives. However, a sufficiently large DM-electron cross section can prevent DM from reaching the solar core, where hotter electrons provide even more substantial velocity boosts.

\paragraph{\label{sec:data}Datasets\textemdash}
We briefly describe the relevant features of the XENON1T and XENONnT datasets used in this analysis, further details are provided in Refs.~\cite{PhysRevLett.123.251801} and~\cite{PhysRevLett.129.161805} respectively.

\paragraph{\label{subsec:1ts2only}XENON1T S2-only dataset: }
Both S1 and S2 signals in a liquid xenon TPC are measured in photoelectrons (PEs), with S2 signals typically 2 orders of magnitude larger in size than S1 signals. DM searches that require coincidence between S1 and S2 signals for event reconstruction have limited sensitivity to DM masses above approximately $3\, \text{GeV/}c^2$, as lower-mass DM interactions often produce S1 signals below the detection threshold~\cite{XENON:2025vwd,LZ:2024zvo,PandaX:2024qfu,XENON:2024hup}. Removing the requirement that both S1 and S2 signals be detected and selecting events based on the S2 signal alone lowers the detection threshold further, extending sensitivity to lower-mass DM. This selection does not exclude events with a detectable S1 signal.

A total of 30\% of the dataset, corresponding to 77.5 live days from XENON1T science run 1~\cite{XENON:2018voc}, is used to optimize event-selection criteria. The remaining 70\%, comprising 180.7 live days, is reserved for the final science search. The region of interest (ROI) is defined in terms of the S2 signal size, limited to S2 peaks within the range of $[150, 3000]\, \text{PE}$, which corresponds to $0.2$ to $3.7\, \text{keV}$ in ER energies~\cite{PhysRevLett.123.251801}. The region below $150\, \text{PE}$ is excluded due to contamination from pileup and single-electron signals following large S2s, which remain poorly understood~\cite{XENON:2021qze,PhysRevLett.123.251801,XENON:2024znc}. We impose an upper bound of $3000\, \text{PE}$, above which the ER background consists of a mixture of events with only S2 signals and those with both S1 and S2 signals. No validated background model exists for this mixture.

\begin{figure}[ht]
    \centering
    \includegraphics[width=\columnwidth]{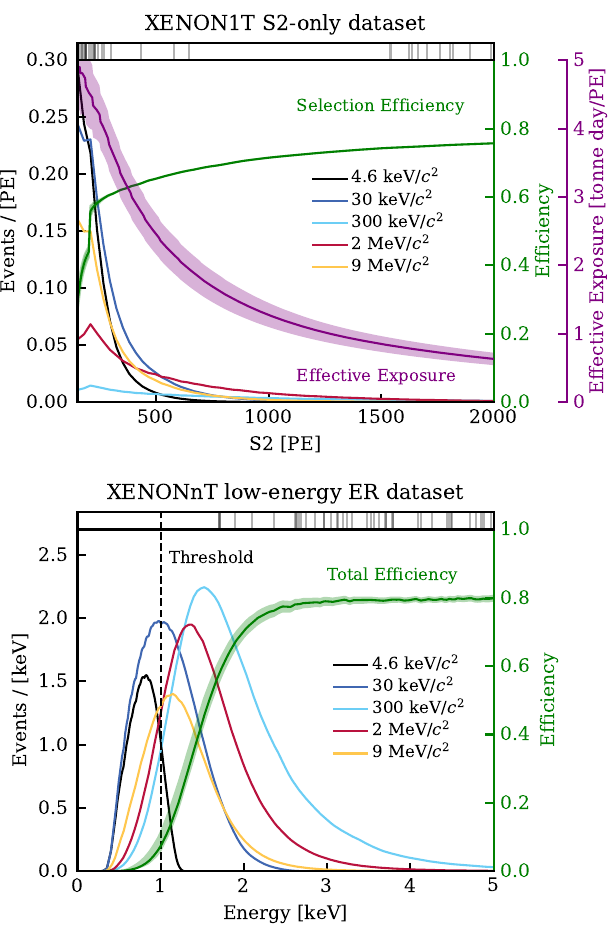}
    \caption{Top: SRDM differential rates as a function of S2 area in the XENON1T detector evaluated at the 90\% CL upper limits on the DM-electron cross section for the various SRDM masses (colors). The spectra are displayed up to 2000 PE to highlight the steeply falling signal spectra. Short vertical gray ticks in the narrow top panel indicate events in the XENON1T S2-only science search dataset. Green line (shaded band) shows the event-selection efficiencies ($1\sigma$ uncertainty). Purple line (shaded band) shows the effective remaining exposure ($1\sigma$ uncertainty) after selections. Bottom: SRDM differential rates as a function of reconstructed energy in the XENONnT detector evaluated at the 90\% CL confidence interval on the DM-electron cross section for the various SRDM masses (colors). Short vertical gray ticks in the narrow top panel indicate events in the XENONnT low-energy ER dataset. Dashed vertical line indicates the 1 keV threshold. Green line (shaded band) shows the combined detection and event-selection efficiencies ($1\sigma$ uncertainty) of the dataset.}
    \label{fig2:signal_spectra}
\end{figure}

We apply the same event-selection criteria as in Ref.~\cite{PhysRevLett.123.251801} to the science search data, and 45 events remain in the ROI of $\text{S2}\in[150,3000]\, \text{PE}$. The S2s of these events are indicated by the short vertical gray ticks in the upper panel of the top subplot in Fig.~\ref{fig2:signal_spectra}.

To obtain the SRDM differential event rate as a function of S2, referred to as the S2 spectrum, we apply an energy threshold of $50\, \text{eV}$ to the deposited energy before transforming the differential rate from deposited energy to S2 size using the detector response matrix provided in the data release~\cite{1t_s2only_datarelease}. The $50\, \text{eV}$ threshold neglects upward fluctuations from subthreshold deposits into the analysis region but is conservative and has a negligible effect on the results. The response matrix incorporates detector fluctuations, reconstruction effects, and selection efficiencies. The resulting S2 spectra are steeply falling. Since the S2 spectra depend on both DM mass and DM-electron cross section, we show only those evaluated at the 90\% CL upper limit on the DM-electron cross section obtained from the XENON1T S2-only dataset, as displayed in the top subplot of Fig.~\ref{fig2:signal_spectra}. The step around 200 PE in the selection efficiency arises from a selection applied only for $\text{S2}<200\, \text{PE}$ to suppress single-electron pileup, and is fully propagated into the signal spectra.

\paragraph{\label{subsec:ntlower}XENONnT low-energy ER dataset: }
Events in the XENONnT low-energy ER dataset are required to have a valid pairing of S1-S2. In addition, S1 candidates must satisfy a threefold tight coincidence requirement, defined as signals in at least three PMTs within a $\pm 50$ ns window around the time of the maximum waveform amplitude~\cite{PhysRevLett.129.161805}. While these requirements result in a higher detector threshold of $\sim1\, \text{keV}$ compared to the XENON1T S2-only dataset, the XENONnT low-energy ER dataset is substantially cleaner as the novel subsystems in XENONnT reduced the ER backgrounds by approximately a factor of 5 relative to XENON1T~\cite{PhysRevLett.129.161805}. In addition, having both S1 and S2 signals provides sufficient information to suppress backgrounds that dominate in ionization-only analyses.

The first science run of XENONnT, conducted from 6 July to 10 November 2021, yielded a live time of 97.1 days. With a fiducial mass of $(4.37 \pm 0.14)$ metric tons, the low-energy ER dataset yields an exposure of $1.16$ metric ton-years~\cite{PhysRevLett.129.161805}. The analysis is performed in reconstructed energy with an ROI spanning $1-140\, \text{keV}$. Although the SRDM signal diminishes above $5\, \text{keV}$ (see Fig.~\ref{fig2:signal_spectra}, bottom), the ROI is extended to $140\, \text{keV}$ to constrain the flat background components more effectively.

We apply the same event-selection criteria as in Ref.~\cite{PhysRevLett.129.161805}. The remaining 3658 events are indicated by short vertical gray ticks at their reconstructed energies in the upper panel of the bottom subplot in Fig.~\ref{fig2:signal_spectra}.

To obtain the SRDM differential rate in reconstructed energy, we convolve the true recoil spectrum with the detector response, incorporating both energy resolution, modeled using a skewed Gaussian~\cite{PhysRevLett.129.161805,Szydagis:2021hfh,LUX:2020car}, and the total efficiency, which is primarily determined by the S1 threefold coincidence requirement and waveform-dependent reconstruction effects~\cite{nt_analysis_paper1}.

Accidental coincidence (AC) events, formed by the random pairing of uncorrelated S1 and S2 signals into spurious events, contribute to the background and are removed by the AC cut. We also apply selection acceptances and data quality cuts, including a $500\, \text{PE}$ S2 threshold to further suppress AC background events. The resulting signal spectra, which incorporate all detector and selection effects, are shown in the bottom subplot of Fig.~\ref{fig2:signal_spectra}.

The signal expectation peaks for a $300\, \text{keV/}c^2$ SRDM, as shown by the light blue curve in the bottom subplot of Fig.~\ref{fig2:signal_spectra}, where the solar upscattering is most efficient for DM masses comparable to the electron mass. DM models with lower masses are suppressed by the detection efficiency and selection acceptances of the XENONnT low-energy ER dataset, while those with higher masses are suppressed by the reduced local DM number density.

\begin{figure}[hb]
    \centering
    \includegraphics[width=\columnwidth]{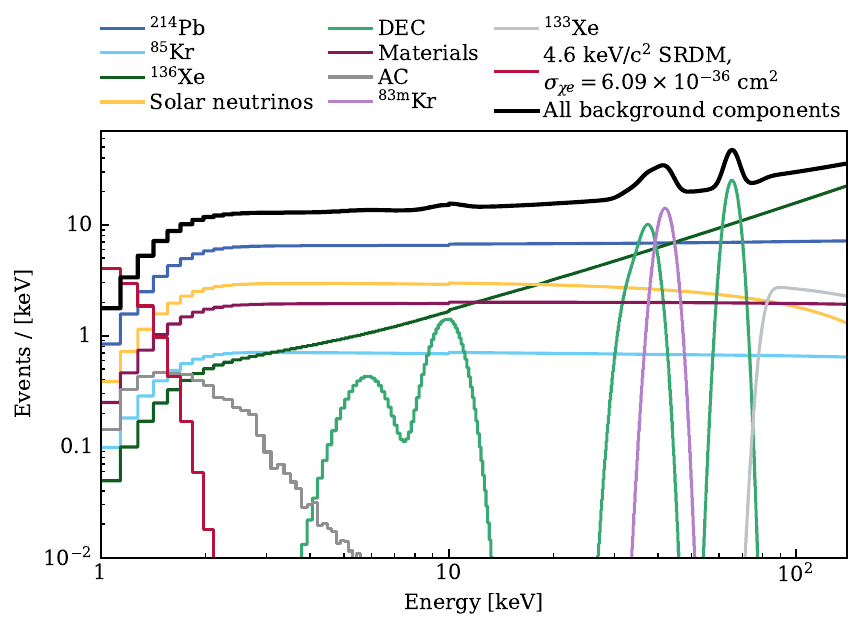}
    \caption{Event expectation for the various background components in the XENONnT low-energy ER dataset. The spectrum of a $4.6\, \text{keV/}c^2$ SRDM signal with DM-electron cross section of $6.09\times10^{-36}\, \text{cm}^{2}$ (unconstrained 90\% CL confidence interval) is also shown for reference. The discontinuity near 10 keV in the event expectation arises from the blinded weakly interacting massive particle search region.}
    \label{fig3:bg}
\end{figure}

We adopt the same background model and associated constraints for the XENONnT low-energy ER dataset as in Ref.~\cite{PhysRevLett.129.161805}. The energy spectra of the various background components included in this model are shown in Fig.~\ref{fig3:bg}.
The dominant contribution to the low-energy background arises from the $\beta$ decay of $^{214}\text{Pb}$, a progeny of the primordial $^{238}\text{U}$ decay chain, followed by ERs induced by solar neutrino scattering. Additional background components include AC events, and $\gamma$ rays from radioactive decays in materials surrounding the detector and the $\beta$ decay of anthropogenic $^{85}\text{Kr}$. Intrinsic radioactivity within the liquid xenon target introduces contributions from the two-neutrino double-$\beta$ decay of $^{136}\text{Xe}$ and the double-electron capture (DEC) of $^{124}\text{Xe}$. Residual contamination from prior calibration activities includes $^{83\text{m}}\text{Kr}$ and $^{133}\text{Xe}$, the latter produced through neutron activation during $^{241}\text{AmBe}$ calibration campaigns conducted a few months before the start of the science run.

\paragraph{\label{sec:stats}Statistical analysis\textemdash}
Because of the lack of a complete background model for the XENON1T S2-only dataset, we employ Yellin’s optimum interval method with the $P_{\text{max}}$ test statistic~\cite{PhysRevD.66.032005} to set upper limits on the DM-electron scattering cross section, using the S2 area as the observable. The statistical trial factor associated with testing multiple intervals is taken into account using Monte Carlo simulations~\cite{tan_yellinpmax_2024}.

In contrast, for the XENONnT low-energy ER dataset, where a full background model is available, we employ an unbinned likelihood analysis over the reconstructed energy range from 1 to $140\, \text{keV}$. As in Ref.~\cite{PhysRevLett.129.161805}, the confidence intervals on the DM-electron scattering cross section for the various DM mass hypotheses are derived using the Feldman-Cousins unified interval construction with the profile likelihood ratio as the test statistic, assuming its asymptotic distribution of a $\chi^2$ distribution with 1 degree of freedom~\cite{Feldman:1997qc,Baxter:2021pqo,Wilks:1938dza}. To account for systematic uncertainties in the detection efficiency near the energy threshold, specifically for energies below $5\, \text{keV}$, we introduce a shape nuisance parameter that allows the efficiency curve to vary.

For DM models in which the DM-electron cross section affects only the total event rate, the confidence interval on the rate multiplier is computed for a fixed reference DM-electron cross section and subsequently translated into a confidence interval on the DM-electron cross section based on the reference DM-electron cross section. In contrast, for the SRDM signal model, the DM-electron cross section influences both the spectral shape and normalization. Therefore, the DM-electron cross section is treated directly as the parameter of interest. The corresponding expected signal spectrum is obtained by interpolation between spectra generated at discrete cross section values, while the rate multiplier is fixed to unity such that the event rate is determined entirely by the cross section.

\begin{figure}[ht]
    \centering
    \includegraphics[width=\columnwidth]{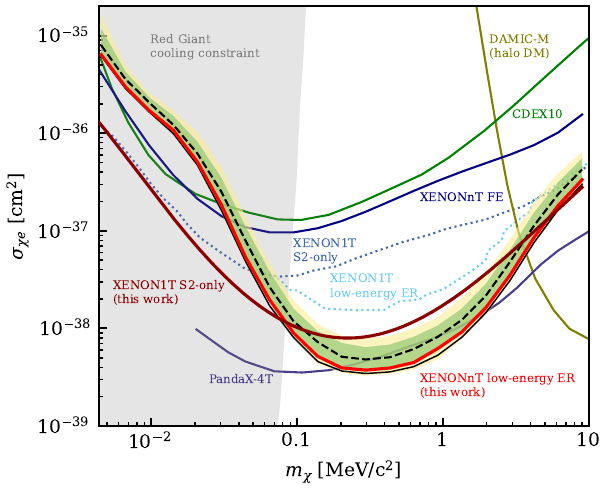}
    \caption{The dark red line shows the 90\% CL upper limits on the DM-electron cross section using the XENON1T S2-only dataset. Because of the lack of a full background model for this dataset, Yellin’s optimum interval method is used~\cite{PhysRevD.66.032005} and no sensitivity band can be constructed. The red line shows the 90\% CL confidence intervals using the XENONnT low-energy ER dataset after applying the power constraint. Thin solid black line shows the 90\% CL confidence intervals computed using only the XENONnT low-energy ER dataset without power constraint, while the dashed black line, along with the green and yellow bands, indicate the median expected sensitivity and the corresponding $1\sigma$ and $2\sigma$ bands respectively. We also show results from XENONnT FE analysis~\cite{XENON:2024znc}, PandaX~\cite{PandaX:2024syk}, CDEX10~\cite{PhysRevLett.132.171001}, and theoretical recasts of limits using XENON1T datasets~\cite{Emken:2021lgc,PhysRevD.104.103026} in dotted. The gray shaded region at low DM mass indicates the parameter space excluded by stellar cooling constraints from red giants~\cite{srdm_rg_Chang_2021}.}
    \label{fig4:results}
\end{figure}

\paragraph{\label{sec:results}Results\textemdash}
Figure~\ref{fig4:results} shows the 90\% CL confidence intervals on the DM-electron cross section obtained from the XENON1T S2-only and XENONnT low-energy ER datasets. Each dataset is analyzed independently, using its respective statistical inference method, and the results are presented together for a coherent view of the constraints on the DM-electron cross section for the various DM masses. Existing constraints include those from the XENONnT few electrons (FE) analysis~\cite{XENON:2024znc} which provides the lowest accessible threshold for XENONnT analyses considered here, other direct DM detection experiments~\cite{PandaX:2024syk,PhysRevLett.132.171001}, and theoretical recasts of XENON1T data~\cite{Emken:2021lgc,PhysRevD.104.103026}. 

In the previous theoretical recast using XENON1T S2-only data~\cite{Emken:2021lgc}, the electron-yield distribution was evaluated with a truncated summation over electron multiplicities, leading to an underestimated signal expectation. In this Letter, we use the emission model from a Monte Carlo simulation of the detector response, which includes contributions from all relevant multiplicities, resulting in a larger predicted signal and correspondingly stronger upper limits. In addition, Ref.~\cite{Emken:2021lgc} considered S2 data only up to 350 PE, whereas we use the full range up to 3000 PE. The optimum interval method identifies a gap around 646--1541 PE, as shown in the narrow panel above the top plot in Fig.~\ref{fig2:signal_spectra}, providing additional constraining power and contributing to the stronger limits.

No sensitivity band is shown for the XENON1T result due to the lack of a full background model. For the XENONnT dataset, the median sensitivity along with the 1$\sigma$ and 2$\sigma$ bands are shown. The power-constrained 90\% CL confidence interval is obtained by restricting any underfluctuation to at most $-1\sigma$~\cite{Baxter:2021pqo}. Without the power constraint, the 90\% CL confidence intervals obtained from the XENONnT low-energy ER dataset lie between $-2$ and $-1\sigma$ of the sensitivity band for all mass hypotheses. A goodness-of-fit test of the background-only model against the observed data yields a $p$-value of 0.39, indicating no evidence for significant mismodeling of the background. All nuisance parameters are consistent with their external constraints, with no deviation exceeding $1\sigma$.

Using the XENONnT low-energy ER dataset, we set the most stringent confidence intervals on the DM-electron scattering cross section in the mass range from $0.2$ to $2\, \text{MeV/}c^2$. In addition, the 90\% CL upper limits derived from the XENON1T S2-only dataset exclude previously unconstrained DM-electron cross section in the lower-mass range of $4.6$ to $20\, \text{keV/}c^2$. Although this region is already in tension with stellar cooling bounds from red giant (RG) observations~\cite{srdm_rg_Chang_2021}, we include it here as the constraints from RG cooling and direct detection experiments are complementary and rely on different assumptions and systematics.

\paragraph{\label{sec:summary_outlook}Summary and outlook\textemdash}
Sub-GeV dark matter in the SHM is typically too light to yield detectable signals in liquid xenon TPCs. However, upscattering of these DM particles from a variety of astrophysical objects can boost their velocity, making detection possible. In particular, the SRDM model considers DM scattering off electrons in the Sun, extending the sensitivity of liquid xenon TPCs to the keV--MeV mass regime.

In this Letter, we present novel 90\% CL confidence intervals on the DM-electron scattering cross section. In the mass range between $4.6$ and $20\, \text{keV/}c^2$, and between $0.2$ and $2\, \text{MeV/}c^2$, we exclude previously unconstrained parameter space, reaching a minimum of $3.41\times10^{-39}\, \text{cm}^2$ for a mass of $0.3\, \text{MeV/}c^2$ at 90\% confidence level by considering SRDM models interacting via a heavy mediator. This extension of sensitivity to lower masses is achieved by considering the upscattering of DM in the Sun, without assuming any new interactions.

Future developments will focus on improving the efficiency of the Monte Carlo calculation of the solar reflected DM flux at very low DM-electron cross sections, where scattering becomes rare and event generation becomes computationally expensive. The XENONnT experiment continues to collect data under improved detector conditions, including ongoing efforts to suppress backgrounds from photoionizing impurities, which are expected to lower the accidental coincidence background rate and enhance sensitivity to sub-GeV dark matter.

\vspace{1em}
\paragraph{\label{sec:acknowledgements}Acknowledgments\textemdash}
P.L.T. thanks Timon Emken for the useful cross-checks and discussions regarding the SRDM model.

We gratefully acknowledge support from the National Science Foundation, Swiss National Science Foundation, German Ministry for Education and Research, Max Planck Gesellschaft, Deutsche Forschungsgemeinschaft, Helmholtz Association, Dutch Research Council (NWO), Fundacao para a Ciencia e Tecnologia, Weizmann Institute of Science, Binational Science Foundation, Région des Pays de la Loire, Knut and Alice Wallenberg Foundation, Kavli Foundation, JSPS Kakenhi, JST FOREST Program, and ERAN in Japan, Tsinghua University Initiative Scientific Research Program, National Natural Science Foundation of China, DIM-ACAV+ Région Ile-de-France, and Istituto Nazionale di Fisica Nucleare. This project has received funding and support from the European Union’s Horizon 2020 and Horizon Europe research and innovation programs under the Marie Skłodowska-Curie Grant Agreements No. 860881-HIDDeN and No. 101081465-AUFRANDE. We gratefully acknowledge support for providing computing and data-processing resources of the Open Science Pool and the European Grid Initiative, at the following computing centers: the CNRS/IN2P3 (Lyon, France), the Dutch national e-infrastructure with the support of SURF Cooperative, the Nikhef Data-Processing Facility (Amsterdam, Netherlands), the INFN-CNAF (Bologna, Italy), the San Diego Supercomputer Center (San Diego, U.S.) and the Enrico Fermi Institute (Chicago, U.S.). We acknowledge the support of the Research Computing Center (RCC) at The University of Chicago for providing computing resources for data analysis. We thank the INFN Laboratori Nazionali del Gran Sasso for hosting and supporting the XENON project.

\paragraph{\label{sec:data_availability}Data availability\textemdash}
In this analysis, we use the data and associated resources from the XENON1T S2-only data release~\cite{PhysRevLett.123.251801,1t_s2only_datarelease}. The XENONnT low-energy electronic recoil data is available at Zenodo~\cite{nT_sr0lower_datarelease}.

\bibliography{biblio.bib}
\end{document}